\documentclass[seceq]{ptptex}
\usepackage{graphicx} 
\usepackage{latexsym}
\notypesetlogo
\begin{document}
\begin{flushright}
RUP-17-9 \\
\end{flushright}
\vspace{10mm}
\begin{center}
\Large{ \bf Quark mass function in Minkowski space }
\vspace{15mm}

\large{  Hidekazu {\sc Tanaka} and Shuji {\sc Sasagawa} \\
Department of Physics, Rikkyo University, 
           Nishi-ikebukuro, Toshima-ku Tokyo, Japan, 171\\ }
 \end{center}

\begin{center}

\vspace{25mm}

{\Large ABSTRACT}
 \end{center}
        
\vspace{10mm}
\def\proj{{\bf P}}  
\def\slsh#1{{#1}{\kern-6pt}/{\kern1pt}}  



        
We investigate the properties of quark mass functions in quantum chromodynamics  calculated by the Schwinger-Dyson equation in the strong coupling region, in which the loop integration is performed in Minkowski space. The calculated results are compared with those obtained by integration in Euclidean space. 

\newpage

\section{Introduction}

The Schwinger-Dyson equation (SDE) [1,2]  is one of methods to evaluate nonperturbative phenomena, such as chiral phase transition.  So far, many works for chiral symmetry breaking have been done with the SDE in momentum representation, in which a one-loop contribution is integrated over Euclidean space.  

Some calculations of fermion mass functions with the SDE have been done in Minkowski space.  In Ref.[3], spectral representation for Green functions is assumed, in which the mass functions are calculated in Lorentz-invariant form.  In Ref.[4], explicit one-loop contributions of the mass function have been calculated. However the mass function is evaluated only one iteration from a constant initial mass as an input. 

 It has been pointed out that the spectral functions for the gluon calculated by the Lattice simulation in Euclidean space, which is numerically continued to Minkowski space becomes negative in in some range.[5] Similar behavior has been found by the generalized perturbative calculations.[6]

 Analytic continuation from Euclidean space to Minkowski space is valid in perturbative calculation if pole positions in complex plane of energy are known. However it is not trivial in the nonperturbative region.

So far, the structure of the fermion mass function in the strong coupling region in the entire range of energy and momentum space has not been fully studied in Minkowski space, even at zero temperature.

In the previous paper[7], we formulated the SDE for quantum electrodynamics (QED) in which the momentum integration is performed in Minkowski space without the instantaneous exchange approximation (IEA) [8,9].

In this paper, we apply our previous method to calculate the quark mass function in quantum chromodynamics (QCD) with the SDE in Minkowski space at zero temperature. 

In Sect. 2, we present the SDE for QCD in Minkowski space. In Sect. 3, some numerical results are shown and calculated results are compared with those obtained by the SDE in Euclidean space.   Section 4 is devoted to a summary and some comments. 

\section{The SDE for the quark mass function}

We calculate a quark self-energy $\Sigma(P)$ in QCD in 4-dimensions, which is given by
\begin{eqnarray}
-i\Sigma(P)=\int{d^4Q \over (2\pi)^4}(ig_s)^2C_F\gamma^{\mu}iS(Q)\Gamma^{\nu}iD_{\mu\nu}(K),\end{eqnarray}
where $S(Q)$ and $D_{\mu\nu}(K)$ are propagators of a quark with momentum $Q=(q_0,{\bf q})$ and a gluon with momentum $K=P-Q=(k_0,{\bf k})$, respectively, 
Here, $P=(p_0,{\bf p})$ is a external momentum of the quark. 
The strong coupling constant and the color factor are denoted by $g_s$ and $C_F=4/3$, respectively.

The  quark propagator is given by
\begin{eqnarray}
iS(Q) = {iZ \over \slsh{Q}-m_0-\Sigma(Q)+i\varepsilon}={i \over A(Q)\slsh{Q}-B(Q) +i\varepsilon },
\end{eqnarray}
where $m_0$ is a bare quark mass.

In this paper, we calculate a mass function in the Landau gauge, in which the wave-function renormalization constant is $Z=1$  in one-loop order of perturbation. Therefore, we calculate the self-energy given in Eq.(2$\cdot$1) with $A=1$, $M_M=B=m_0+Tr[\Sigma]/4$, and the quark-gluon vertex with $\Gamma_{\mu}=\gamma_{\mu}$. Here, the gluon propagator is given as 
\begin{eqnarray}
iD_{\mu\nu}(K)=\left(-g_{\mu\nu}+{K_{\mu}K_{\nu} \over K^2}\right){i \over K^2+i\epsilon}
\end{eqnarray}
in the Landau gauge.  In this paper, we neglect an effective gluon mass. Integrating over the azimuthal angle of the momentum ${\bf q}$, the mass function is given by 
\begin{eqnarray}
M_{\rm M}(p_0,p) =m_0 -{3iC_F \over 2\pi^2}\int^{\Lambda_0}_{-\Lambda_0}dq_0 \int^{\Lambda}_{\delta} dq {q \over p} \alpha_s [M_{\rm M}IJ](q_0,q) 
\end{eqnarray}
with $p=|{\bf p}|$,$q=|{\bf q}|$ and $\alpha_s=g_s^2/(4\pi)$. Here,$I$ and $J$ are given by
\begin{eqnarray}
I= {1 \over Q^2-M_{\rm M}^2+i\varepsilon}
\end{eqnarray}
and
\begin{eqnarray}
J=\int^{\eta_+}_{\eta_-}dk{k \over K^2+i\varepsilon},
\end{eqnarray}
respectively, with $\eta_{\pm}=|p\pm q|$ and $k=|{\bf k}|$.

As implemented in the previous work for QED [7], we approximate the integration over $q_0$ as
\begin{eqnarray}
M_{\rm M}(p_0,p) \simeq m_0-{3iC_F \over 2\pi^2} \int^{\Lambda}_{\delta} dq {q \over p}\sum_{l=1}^{N-1}\alpha_s <[M_{\rm M}J](q)>_l I(q_0^{(l+1)},q_0^{(l)}).
\end{eqnarray} 
with
\begin{eqnarray}
I(q_0^{(l+1)},q_0^{(l)}) = \int^{q_0^{(l+1)}}_{q_0^{(l)}}{dq_0 \over Q^2-M_{\rm M}^2+i\varepsilon},
\end{eqnarray}
where, $<X>_l$ denotes an average of  $X(q_0^{(l+1)})$ and $X(q_0^{(l)})$ as $ <X>_l=[X(q_0^{(l+1)})+X(q_0^{(l)})]/2$.

The explicit expressions are summarized in Appendix A.

\section{Numerical results}

In this section, some numerical results are presented. We solve the SDE presented in Eq. (2$\cdot$1) by a recursion method starting from a constant mass. \footnote{Initial input parameters are $M_{\rm R}=\Lambda_{QCD}$ and $M_{\rm I}=0$ with $\Lambda_0=\Lambda=20\Lambda_{\rm QCD}$ and $\delta=0.2\Lambda_{\rm QCD}$, where we define $M_{\rm M}= M_{\rm R}+i M_{\rm I}$. 
We set $\Lambda_{\rm QCD}=0.5GeV$ with $\varepsilon=10^{-6}$.}

For each iteration, we calculate the quark mass function normalized as 
\begin{eqnarray}
M^{(n+1)}_{\rm M}(P^2) = m(\mu^2)+M_{\rm M}^{(n)}(P^2)-M_{\rm M}^{(n)}(\mu^2), 
\end{eqnarray}
where $n$ denotes the number of iterations. Here, the mass function is normalized by a current quark mass at large $\mu^2$, in which perturbative calculations are  reliable. In the iteration, the mass function $M_{\rm M}(p_0,p)$ in integrand of   Eq.(2.4) is replaced by the renormalized one obtained by the previous iteration. Here, $m(\mu^2)$ is a renormalized mass at a renormalization scale $\mu$.\footnote{We take $m(\mu^2)=3{\rm MeV}$ at $\mu=20\Lambda_{QCD}$.  }

 Here, $M_{\rm M}(P^2)$ is defined as 
\begin{eqnarray}
M_{\rm M}(P^2) = M_{\rm M}(p_0,\delta)\theta(|p_0|-\delta) 
\end{eqnarray}
for $P^2>0$, and 
\begin{eqnarray}
M_{\rm M}(P^2) = M_{\rm M}(\delta,p) \theta(p-\delta)
\end{eqnarray}
for $P^2<0$, respectively, where $P^2=p_0^2-p^2$.

The mass function can be written as an absolute value $|M_M(P^2)|$ and a phase factor  $\exp(i\Phi(P^2))$ as 
\begin{eqnarray}
M_{\rm M}(P^2)=|M_{\rm M}(P^2)|\exp(i\Phi(P^2)).
\end{eqnarray}
In Fig.1, the convergence property of the mass function $|M_{\rm M}|$ integrated over $|P^2|$ as
\begin{eqnarray}
<|M_{\rm M}(P^2)|>=\int_{\delta^2}^{\Lambda^2} d|P^2||M_{\rm M}(P^2)|
\end{eqnarray}
are presented.
 The solid line and the $+$ symbols represent $<|M_{\rm M}|>$ integrated over $P^2>0$ and $P^2<0$, respectively. The horizontal axis denotes the number of iterations. The dash-dotted line represents the calculated results for $<M_{\rm E}>$, where the mass function in Euclidean space $M_{\rm E} (P_{\rm E}^2)$ is given by [10],[11]
\begin{eqnarray}
M_{\rm E}(x) = {3C_F \over 4\pi}\int^{\Lambda^2}_{\delta^2} dy\alpha_s{2y \over x+y+\sqrt{(x-y)^2}}{M_{\rm E}(y) \over y+M_{\rm E}^2(y) },
\end{eqnarray}
which is renormalized as 
\begin{eqnarray}
M^{(n+1)}_{\rm E}(P^2_{\rm E}) = m(\mu^2)+M_{\rm E}^{(n)}(P^2_E)-M_{\rm E}^{(n)}(\mu^2). 
\end{eqnarray}
As shown in Fig.1, the mass functions integrated over the momentum for the three cases rapidly converge.

\begin{figure}
\centerline{\includegraphics[width=10cm]{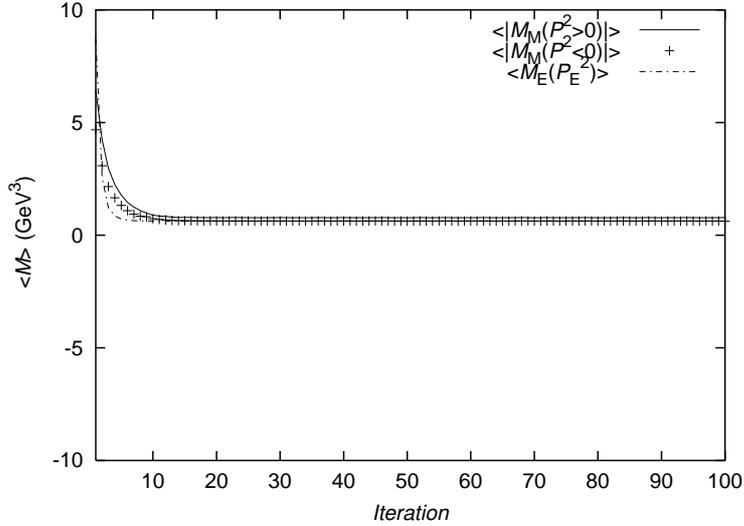}}
\caption{The solid line and the $+$ symbols represent the convergence property of $<M_{\rm M}>$ integrated over $P^2>0$ and $P^2<0$ in Minkowski space, respectively. The dash-dotted line denotes calculated results for $<M_{\rm E}>$ in Euclidean space. The horizontal axis represents the number of iterations. }
\end{figure}

In Fig.2, the dependences on $P^2$ for $<M_i> ~~(i={\rm M,E})$  are presented.
The three cases give similar $|P^2|$ dependences, though the mass function in time-like momentum has an imaginary part.

In Fig.3, $P^2$ dependences of $M_{\rm R}$ and $M_{\rm I}$ are presented.
From Eq.(2$\cdot$2), we define a spectral function for the quark mass term denoted by $\rho_{\rm M}(P^2)$ as
\begin{eqnarray}
{1 \over 4}Tr[S(P)]=S_{\rm M}(P^2)={M_{\rm M}(P^2) \over P^2-M^2_{\rm M}(P^2)+i\varepsilon}=\int dQ^2{\rho_{\rm M}(Q^2) \over P^2-Q^2+i\varepsilon},
\end{eqnarray}
 which gives 
\begin{eqnarray}
\rho_{\rm M}(P^2)=-{1 \over \pi}ImS_{\rm M}(P^2)=- {1 \over \pi}{M_{\rm I}(P^2+|M_{\rm M}|^2)-\varepsilon M_{\rm R} \over (P^2-(M^2)_{\rm R})^2+((M^2)_{\rm I}-\varepsilon)^2}. 
\end{eqnarray}
As shown in Fig.3, 
$M_{\rm I}$ for $P^2>0$  becomes positive in some regions of $P^2$, such as $P^2>0.3{\rm GeV}^2$, in which $\rho_{\rm M}$ becomes negative value. It may be interesting to compare our result with the spectral function for the quark obtained in Ref.[6] in Minkowski space.

\begin{figure}
\centerline{\includegraphics[width=10cm]{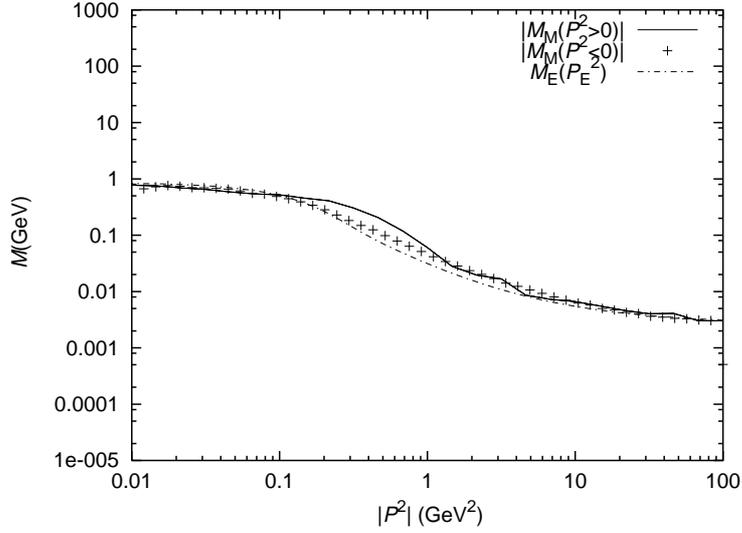}}
\caption{ The solid line and the $+$ symbols represent the $P^2$ dependence of $|M_{\rm M}|$ for $P^2>0$ and $P^2<0$, respectively. The dash-dotted line denotes the result for $P_{\rm E}^2$ dependence of $M_{\rm E}$ in Euclidean space.}
\end{figure}

\begin{figure}
\centerline{\includegraphics[width=10cm]{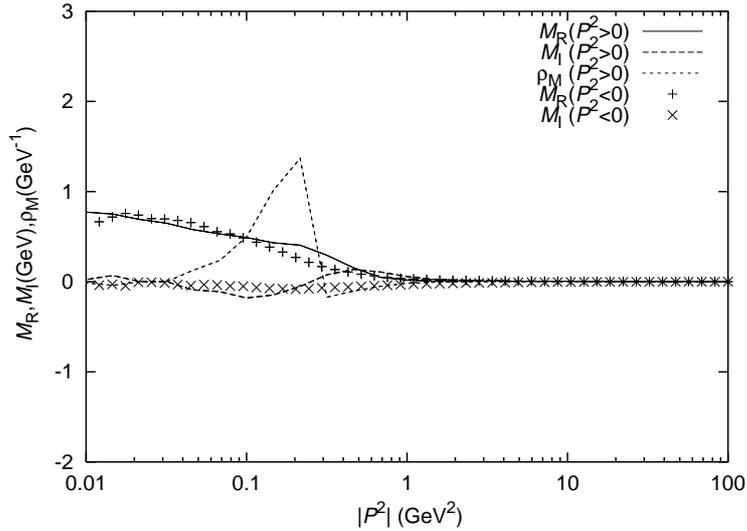}}
\caption{The solid line and the dashed line denote the $P^2$ dependences of $M_{\rm R}$ and $M_{\rm I}$ for $P^2>0$, respectively. $M_{\rm R}$ and $M_{\rm I}$ for $P^2<0$ are represented by the $+$ and the $\times$ symbols, respectively.  The dotted line denotes the $P^2$ dependence of the spectral function $\rho_{\rm M}$ defined in Eq.(3$\cdot$9) for $P^2>0$.}
\end{figure}

\section{Summary and Comments}

In this paper, we studied a quark mass function solved by the Schwinger-Dyson equation (SDE) in Minkowski space for QCD. 

Evaluation of the mass functions in Minkowski space allow us to study an imaginary part of the mass and energy for the massive fermion states.  

We examined the properties of the quark mass function in time-like momentum $P^2>0$ as well as that in space-like momentum $P^2<0$, where $P^2$ denotes a squared four-momentum of the quark.  

Furthermore, we also compared our results with the mass function calculated in Euclidean space.
 
We found that the three cases give similar $|P^2|$ dependences, though the mass function with time-like momentum has an imaginary part.

We also studied a behavior of the spectral function for the quark mass term.
We found that there seems to exist negative spectral function in some momentum regions for $P^2>0$ as pointed out in Refs.[5,6], in which the imaginary part of the mass function becomes positive value.

It may be expected that the SDE has multiple solutions in numerical calculations.  
Further studies are needed for solutions of the mass function of quark obtained by the SDE in  Minkowski space in the strong coupling region. 

In this paper, we examined the qualitative features of the quark mass function in Minkowski space. 
In order to reproduce physical quantities, such as the pion decay constant, or to fit the results obtained by SDE with the numerical data by Lattice simulations, we need  to fine-tune the parameter $\Lambda_{\rm QCD}$.

In future works, we shall extend our method to finite temperature and density with the real time formalism.

\section*{Acknowledgements}

This work was  supported by MEXT-Supported Program for the Strategic Research Foundation at Private Universities, 2014-2017.



\vspace{5mm}

\begin{center}
{\Large Appendix A}
\end{center}
\vspace{5mm}

The mass function is given by 
$$
M_{\rm M}(p_0,p) =m_0 -{3iC_F \over 2\pi^2}\int^{\Lambda_0}_{-\Lambda_0}dq_0 \int^{\Lambda}_{\delta} dq {q \over p}[\alpha_sM_{\rm M}IJ](q_0,q), 
$$
where, the propagators $ I$ and $J$ are defined in Eq.(2$\cdot$5) and Eq.(2$\cdot$6), respectively.

As shown in Eq.(2$\cdot$7), we approximate the integration over $q_0$ as
$$
M_{\rm M}(p_0,p) \simeq m_0-{3iC_F \over 2\pi^2} \int^{\Lambda}_{\delta} dq {q \over p}\sum_{l=1}^{N-1}\alpha_s({\bar P}^2,{\bar Q}^2)<[M_{\rm M}J](q)>_l I(q_0^{(l+1)},q_0^{(l)}),
$$ 
where, $<X>_l$ denotes an average of  $X(q_0^{(l+1)})$ and $X(q_0^{(l)})$ as $ <X>_l=[X(q_0^{(l+1)})+X(q_0^{(l)})]/2$.

Here, the strong coupling constant $\alpha_s$ is replaced by the running coupling constant $\alpha_s({\bar P}^2,{\bar Q}^2)= g^2_s({\bar P}^2,{\bar Q}^2)/( 4\pi) $[12], which is defined as
$$
g^2_s({\bar P}^2,{\bar Q}^2)={1 \over \beta_0} \times \left\{ \begin{array}{ll}
     {1 \over t}  & {\rm if}  ~~t_{\rm F} < t \\
     {1 \over t_{\rm F}}+{(t_{\rm F}-t_{\rm C})^2-(t_{\rm C}-t)^2 \over 2t_{\rm F}^2(t_{\rm F}-t_{\rm C})}  & {\rm if} ~~ t_{\rm C} < t < t_{\rm F}    \\
      {1 \over t_{\rm F}}+{(t_{\rm F}-t_{\rm C}) \over 2t_{\rm F}^2}  &  {\rm if} ~~t < t_{\rm C}  \end{array} \right\}
$$
with $\beta_0=(33-2N_{\rm f})/(48\pi^2)$,$t=\log[({\bar P}^2+{\bar Q}^2)/\Lambda^2_{\rm QCD}],t_{\rm F}=0,5$ and $t_{\rm C}=-2$ for $N_{\rm f}$ flavours, where, ${\bar P}^2=p_0^2+p^2$ and ${\bar Q}^2=<q_0>_l^2+q^2$.
\footnote{In order to compare our results in Minkowski space with those obtained in Euclidean space, we implement the QCD coupling constant $\alpha_s$ with Euclidean momenta. Difference between the argument of $\alpha_s$ with the momenta in Minkowski space and that in Euclidean space is a part of higher order contributions to the one-loop approximation. }

For $I,M_{\rm M}$ and $J$,  we separate the real parts $I_{\rm R},M_{\rm R},J_{\rm R} $ and the imaginary parts $I_{\rm I},M_{\rm I},J_{\rm I}$, respectively.

For $J$ in Eq.(2$\cdot$6), we can integrate over $k$ as
$$
 J_{\rm R} = -\int^{\eta_+}_{\eta_-}dk{k(k^2-k_0^2) \over (k^2-k_0^2)^2+\varepsilon^2} =  -{1 \over 4}\log{(\eta_+^2-k_0^2)^2+\varepsilon^2 \over (\eta_-^2-k_0^2)^2+\varepsilon^2} 
$$
 and
$$
 J_{\rm I}= -\int^{\eta_+}_{\eta_-}dk{k\varepsilon \over  (k^2-k_0^2)^2+\varepsilon^2}    
= -{1 \over 2}\left[\arctan{\eta_+^2-k_0^2 \over \varepsilon}-\arctan{\eta_-^2-k_0^2 \over \varepsilon}\right],
$$
respectively, with $\eta_{\pm}=|p\pm q|$ and $k=|{\bf k}|$.

The real  and imaginary parts of the quark propagator $ I(q_0^{(l+1)},q_0^{(l)})$ defined in Eq.(2$\cdot$8) are given by
$$ I_{\rm R}^{(l)}={\rm Re}I(q_0^{(l+1)},q_0^{(l)}) ={\epsilon(<2q_0-{\partial (E^2)_{\rm R}\over\partial q_0}>_l) \over 2<|2q_0-{\partial (E^2)_{\rm R} \over\partial q_0}|>_l } $$
$$
\times \log{[(q_0^{(l+1)})^2-<(E^2)_{\rm R}>_l]^2+<(E^2)_{\rm I}>_l^2 \over [(q_0^{(l)})^2-<(E^2)_{\rm R}>_l]^2+<(E^2)_{\rm I}>_l^2} 
$$
and
$$ I_{\rm I}^{(l)}={\rm Im}I(q_0^{(l+1)},q_0^{(l)})={\epsilon(<2q_0-{\partial (E^2)_{\rm R}\over\partial q_0}>_l)\epsilon(<(E^2)_{\rm I}>_l) \over <|2q_0-{\partial (E^2)_{\rm R} \over\partial q_0}|>_l } $$
$$
 \times \left[\arctan{(q_0^{(l+1)})^2-<(E^2)_{\rm R}>_l \over |<(E^2)_{\rm I}>_l|}-\arctan{(q_0^{(l)})^2-<(E^2)_{\rm R}>_l \over |<(E^2)_{\rm I}>_l|}\right], 
$$
respectively,  where, we define $\epsilon(z)=\theta(z)-\theta(-z)$ with the step function $\theta(z)$. Here, the real and imaginary parts of the squared energy denoted by $(E^2)_{\rm R}$ and $(E^2)_{\rm I}$, respectively, are given as
$$
 (E^2)_{\rm R}=q^2+(M^2)_{\rm R} 
$$
and
$$
 (E^2)_{\rm I}=(M^2)_{\rm I}-\varepsilon,
$$
with 
$$
 (M^2)_{\rm R}={\rm Re}(M_{\rm M}^2)=(M_{\rm R})^2-(M_{\rm I})^2 
$$
and
$$
 (M^2)_{\rm I}={\rm Im}(M^2_{\rm M})=2M_{\rm R}M_{\rm I}.
$$


\end{document}